\documentclass[letterpaper,journal]{IEEEtran}

\usepackage{graphicx,psfrag,dblfloatfix}
\usepackage{amsmath,amssymb,amsfonts,mathrsfs}
\usepackage{color,epstopdf}
\usepackage{algorithmic}
\usepackage{enumerate} 
\usepackage{subfigure}
\usepackage{tikz}
\usetikzlibrary{shapes,arrows}
\usepackage{float}
\usepackage{flushend}
\usepackage{bbold}
\usepackage{dsfont}
\usepackage{mathrsfs}
\usepackage{authblk}

\newtheorem{example}{Example}
\newtheorem{rem}{Remark}
\newtheorem{assume}{Assumption}
\newtheorem{definition}{Definition}
\newtheorem{theorem}{Theorem}

\newcommand{\bbm}{\begin{bmatrix}}
\newcommand{\ebm}{\end{bmatrix}}

\def\qedp{\hspace*{\fill}~{\tiny $\blacksquare$}}
\def\be{\begin{equation}}
\def\ee{\end{equation}}
\def\ba{\begin{array}}
\def\ea{\end{array}}
\def\eqa{\begin{eqnarray}}
\def\eqe{\end{eqnarray}}

\definecolor{darkgreen}{rgb}{0.0, 0.55, 0.0}
\definecolor{amaranth}{rgb}{0.9, 0.17, 0.31}
\usepackage[prependcaption,colorinlistoftodos]{todonotes}

\begin{document}
\title{On data-driven stabilization of  systems with quadratic nonlinearities}

\author[*]{Alessandro Luppi \thanks{This project is conducted under the auspices of the Centre for Data Science and Systems Complexity at the University of Groningen and is supported by a Marie Sk\l{}odowska-Curie COFUND grant, no. 754315.}}

\author[*]{Claudio De Persis}
\author[**]{Pietro Tesi}
\affil[*]{\textit{ENTEG, University of Groningen, 9747 AG Groningen, The
	Netherlands (email: {a.luppi, c.de.persis}@rug.nl).}}
\affil[**]{\textit{DINFO, University of Florence, 50139 Florence, Italy (email:
	pietro.tesi@unifi.it})}
\maketitle

\begin{abstract}
In this paper, we directly design a state feedback controller that stabilizes a class 
of uncertain nonlinear systems solely based on input-state data collected 
from a finite-length experiment.  Necessary and sufficient conditions are derived
to guarantee that the system is absolutely stabilizable and a controller is designed.
Results derived under some relaxed prior information about the system and strengthened 
data assumptions are also discussed. Numerical examples illustrate the method 
with different levels of prior information.
\end{abstract}

\section{Introduction}
Historically, the design of a stabilizing controller relies on the knowledge of a mathematical model to represent the underling system. More recently, the necessity of identifying a model was reconsidered in favor of a direct data driven design, where controllers are directly synthesized from experimental data. Various efforts have been made in this direction and we refer the interested reader to   \cite{hou2013model} for a survey on the topic. A direct design has the advantage that it does not suffer from modeling errors and the overall controller design is simplified by removing one step as there is no intermediate identification step. This can be especially appealing when dealing with nonlinear systems as deriving a reliable model is challenging and prone to errors. One of the main open research questions on data-driven control is how to provide  stabilizing conditions for the learned controller which are provably correct. A promising result comes from behavioral system theory \cite{willems2005note}, where it has been proved that a single data trajectory obtained from one experiment can be used to represent all the input-output trajectories of a linear time-invariant system. Since \cite{coulson2019data}, which highlighted  the relevance of \cite{willems2005note} for the synthesis of data-enabled predictive control, several papers have found  inspiration from 
the paper of Willems et al.

The result has been revisited through classic state-space description in \cite{depersis-tesi2020tac}, where a data driven parametrization of a closed loop system was derived and used to learn feedback controllers for unknown systems from data. 
Using this setting, several recent contributions have further explored the role of data in synthesis problems in multiple areas of control, including: optimal and robust control \cite{depersis-tesi2020tac}, \cite{holicki2020controller}, robust stabilization of $\mathcal{H}_2, \mathcal{H}_{\infty}$ control with noisy data \cite{depersis-tesi2020tac}, \cite{berberich2020robust}, \cite{dpt2021Aut}, \cite{van2020noisy}, \cite{berberich2019combining}, dissipativity proprieties \cite{romer2017determining}, robust set-invariance \cite{Bisoffi2020}, and the $L_2$-gain $\cite{sharf2020sample}$.
The design of a data-driven stabilizing controller for nonlinear systems has been approached using various methods and for different classes of nonlinear systems  \cite{novara2016data}, \cite{kaiser2017data}, \cite{lusch2018deep}, \cite{Bisoffi2020}, \cite{guo2019data}, \cite{dai2020semi}. 

In this paper, we focus on finding a data-driven solution to the problem of absolute stabilizability. Absolute stabilizability is the problem of enforcing the stability of the origin via feedback for a class of systems comprising a linear part and nonlinearities that satisfy a given condition. In our work we impose quadratic restrictions on the nonlinearity. 
The absolute stability problem was originally formulated by A.I. Lurie in \cite{lurie1957some} and solved with a Lyapunov based approach. Later in \cite{popov1958relaxing} V.M. Popov proposed a solution in the frequency domain. These two approaches were later connected by the Kalman-Yakubovich-Popov Lemma \cite{yaku} that relates an analytic property of a square transfer matrix in the frequency domain to a set of algebraic equations involving parameters of a minimal realization in time domain. 

To solve the absolute stabilizability problem in the case in which the system is unknwon, we start from the data driven parametrization in \cite{depersis-tesi2020tac} for a closed loop system with the addition of a nonlinear term. Then we propose a data-dependent Lyapunov-based control design assuming that a finite number of samples measuring the nonlinear term is available. 
One of the advantage of our formulation, compared to a model based one, is that it holds for both continuous-time and discrete-time systems providing a unified analysis and design framework for both classes of systems. We also discuss how our results can be viewed in a frequency domain stability analysis where our main result can be interpreted as a data-dependent feedback Kalman-Yakubovitch Popov Lemma \cite[Section 2.7.4]{fradkov2013nonlinear}. Finally, we discuss how to deal with different levels of prior knowledge 
by strengthening the assumption on the collected data. 

\textbf{Contribution.} Our main contributions are necessary and sufficient conditions for which a solution to the data-driven absolute stabilization problem exists. The proposed conditions can be verified directly from data by means of efficient linear programs. These conditions provide a new data based solution to the problem of absolute stabilizability for a class of uncertain nonlinear systems. 

\textbf{Structure.} The paper is organized as follow. The notation and problem setup is presented in Section \ref{sec:framework}. 
In Section \ref{sec:finite-samples-nonlin}, we present necessary and sufficient conditions under which the data-driven absolute stabilizability problem is solvable and provide the explicit expression for the controller. Finally in Section \ref{sec:relaxing}, we relax some prior  knowledge about the system and  derive new conditions for the problem solution. All the main results are illustrated also with practical examples. Concluding remarks are given in Section \ref{sec:concl}.

\section{Framework and problem formulation} \label{sec:framework}

\tikzstyle{block} = [draw, fill=white, rectangle, 
    minimum height=3em, minimum width=6em]
\tikzstyle{sum} = [draw, fill=white, circle, node distance=1cm]
\tikzstyle{input} = [coordinate]
\tikzstyle{input1} = [coordinate]
\tikzstyle{output} = [coordinate]
\tikzstyle{output2} = [coordinate]
\tikzstyle{pinstyle} = [pin edge={to-,thin,black}]

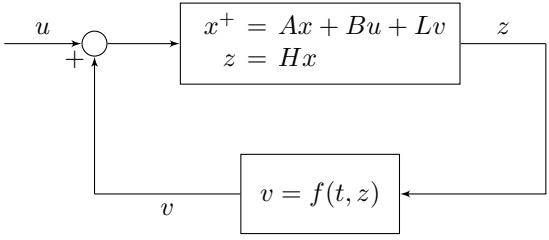
\begin{figure}
\centering
\begin{tikzpicture}[auto, node distance=2cm,>=latex']
    \node [input, name=input] {};
    \node [sum, right of=input] (sum) {};
    \node [input1, left of=sum, node distance=1.2cm] (input1) {};
    \node [block, right of=sum, node distance=3cm] (system) 
    	      {\setlength\arraycolsep{2pt}  $\begin{array}{rcl} 
	      x^+ &=& Ax + Bu + Lv \\ z &=& Hx \end{array}$};
    \node [output, right of=system, node distance=3cm] (output) {};
    \node [block, below of=system] (nonlinearity) 
    {$v=f(t,z)$};
    \node [output2, left of=nonlinearity, node distance=3cm] (output2) {};
    \draw [->] (input1) -- node [name=u] {$u$}(sum);
    \draw [->] (sum) -- node {} (system);
    \draw [-] (system) -- node [name=z] {$z$}(output);
    \draw [->] (output) |- (nonlinearity);
    \draw [-] (nonlinearity) -- node [name=v] {$v$}(output2);
    \draw [->] (output2) -| node[pos=0.99] {$+$} 
        node [near end] {} (sum);
\end{tikzpicture}
\caption{Schematic diagram of system \eqref{nonlinear.system2}.}
\label{fig:diagram}
\end{figure}

We study the stabilization problem of a nonlinear system of the form
\be\label{nonlinear.system}
x^+= Ax + Bu + \hat f(t,x)
\ee
where $x\in \mathbb{R}^n$ is the state, $u\in \mathbb{R}^m$ 
is the control input, and where
$\hat f(t,x):\mathbb{R}\times \mathbb{R}^n\to \mathbb{R}^n$
is a memoryless, possibly time-varying nonlinearity.
The matrices $A,B$ and the map $\hat f$ are unknown. 
If some prior information is available regarding $\hat f$, 
we will model it in the form $\hat f(t, x)=L v$, $v=f(t,Hx)$, 
with $L$ and $H$ known
matrices and 
$f:\mathbb{R}\times\mathbb{R}^p \to \mathbb{R}^q$. 
The system then becomes (see Figure \ref{fig:diagram})
\be\label{nonlinear.system2}
\setlength\arraycolsep{2pt}
\ba{rcl}
x^+ &= & Ax + Bu + L v\\[0.1cm]
z &=& Hx  \\[0.1cm]
v &= & f(t,z)
\ea
\ee 
If prior information on the input matrix $L$ or $H$ is unavailable,
then we set $L=I_n$, $H=I_n$, and $\hat f(t,x)=f(t,x)$. 

Throughout this paper, we will consider certain constraints 
on the admissible nonlinearities. Specifically, we will assume that
the inequality 
\be\label{in.asspt4-z}
\begin{bmatrix}
z\\ f(t,z)
\end{bmatrix}^\top
\begin{bmatrix}
\hat Q& \hat S\\ \hat S^\top & \hat R
\end{bmatrix}
\begin{bmatrix}
z\\  f(t,z)
\end{bmatrix}\ge 0
\ee
holds for all the pairs $(t,z)\in \mathbb{R}\times \mathbb{R}^p$ with $z \in \text{im}\, H$, 
where $\hat Q=\hat Q^\top \in \mathbb{R}^{p\times p}$, 
$\hat S\in \mathbb{R}^{p\times q}$ and $\hat R\prec 0 \in \mathbb{R}^{q\times q}$ 
are known matrices (definite matrices are implicitly defined as 
symmetric matrices). Since $\hat R\prec 0$ the inequality \eqref{in.asspt4-z} implies 
$f(t,0)=0$ for all $t \in \mathbb{R}$. 
In the remainder, we will sometimes ask  the constraint \eqref{in.asspt4-z} 
to be {\it regular}, by which we mean that there exists a pair
$(\overline z, f(\overline t, \overline z))$ such that the inequality 
\eqref{in.asspt4-z} evaluated at $(\overline z, f(\overline t, \overline z))$ 
strictly holds.

For this class of systems, a 
notion of stability widely studied in the literature is the so-called {\it absolute stability}
which is now introduced (\emph{cf.} \cite[Definition 10.2]{khalil1996nonlinear}).
\begin{definition}
System 
\eqref{nonlinear.system2}
is said to be absolutely stabilizable via linear state-feedback 
$u=Kx$ if there exists a matrix $K$ such that the origin of the closed-loop system  
\be\label{nonlinear.system2-fb}
\setlength\arraycolsep{2pt}
\ba{rcl}
x^+ &= & (A+ BK)x + L v\\[0.1cm]
z &=& Hx  \\[0.1cm]
v &= & f(t,z)
\ea
\ee
is globally uniformly asymptotically stable for any function $f$ 
that satisfies the inequality \eqref{in.asspt4-z}. \qedp
\end{definition}

This framework covers a notable class of nonlinear systems that has appeared in several studies. 

\begin{enumerate}
\item {\it (Lipschitz  or norm bounded  nonlinearities)} 
These are nonlinearities that satisfy 
\[
f(t,z)^\top f(t,z) \le \ell^2 z^\top 
z,
\]
where $\ell$ is a bound on the Lipschitz constant, which are 
considered in LMI-based robust stabilization of nonlinear systems \cite{vsiljak2000robust}. 
In the absolute stability theory literature, e.g. \cite[Section 3]{Haddad1994IJRNC}, 
these nonlinearities are referred to as norm-bounded nonlinearities.
In this case \eqref{in.asspt4-z}  holds with 
$\hat Q = \ell^2 I_{p}$, 
$\hat S= 0$ and $\hat R = -I_{q}$.
\item {\it (Bounds on partial gradients)} Large Lipschitz constants might 
result in unfeasible conditions, and for this reason much literature is devoted to deriving 
less conservative Lipschitz characterizations (\cite{Zemouche2013aut} and 
references therein). Here, we recall a result from \cite{jin-lavaei2018cdc} 
that considers the nonlinear time-invariant term  $\hat f(x)$ instead of $Lf(t,z)$.
Under continuous differentiability of $\hat f$, and assuming that bounds 
$\underline{f}_{ij},\overline{f}_{ij}$ on the partial derivatives are known, namely
\[
\underline{f}_{ij} \le \displaystyle\frac{\partial \hat{f}_i}{\partial x_j} \le \overline{f}_{ij}
\]
it holds that $\hat f(x)= L f(z)$, where $L= I_n \otimes \mathds{1}_n^\top$, $H=I_n$
and \eqref{in.asspt4-z} holds with the matrices 
$\hat Q, \hat S$ and $\hat R$ depending on the vectors of bounds $\overline{f},\underline{f}$. 
For instance, in the case $n=2$, the matrix in \eqref{in.asspt4-z} takes the form 

\vspace{-.3cm}
{\small
\be\label{tilde.Q.Lipschitz}
\left[\ba{cc|cccc}
\displaystyle \sum_{i=1}^2 (\overline{c}_{i1}-{c}_{i1}) & 0 & c_{11} & 0 & c_{21} & 0\\
0 & \displaystyle \sum_{i=1}^2 (\overline{c}_{i2}-{c}_{i2}) & 0 & c_{12} & 0 & c_{22}\\
\hline
\star &\star &  -1 & 0 & 0 & 0  \\
\star &\star &  0 & -1 & 0 & 0  \\
\star &\star &  0 & 0 & -1 & 0  \\
\star &\star &  0 & 0 & 0 & -1    \\
\ea
\right]
\ee
}%
where $c_{ij} = (\overline{f}_{ij}+\underline{f}_{ij})/2$ and $\overline{c}_{ij} 
= (\overline{f}_{ij}-\underline{f}_{ij})/2$. 
For additional degree of freedom, one can introduce a vector $\lambda$ of non-negative multipliers,
see \cite{jin-lavaei2018cdc} for details.

\item {\it (Strongly convex functions with Lipschitz gradient)} 
Let $\hat f(x)=\nabla g(x)$, with $\hat f(0)=0$ and 
$g:\mathbb{R}^n\to \mathbb{R}$ a continuously differentiable strongly 
convex function with parameter $m$ and having Lipschitz gradient with 
parameter $\ell$, with $0<m<\ell$. Then condition \eqref{in.asspt4-z} holds with 
$\hat Q=-2m\ell I_n$, $\hat S= (\ell+m) I_n$ and $\hat R= -2 I_n$, 
see \cite[Proposition 5, (3.13d)]{lessard2016analysis}.  

\item ({\it Sector bounded nonlinearities}) 
Another notable case is when
the nonlinear function $f$ satisfy the sector condition 
$(f(t,z)-K_1 z)^\top (K_2 z-f(t,z))\ge 0$ for $(t,z)\in \mathbb{R}\times \mathbb{R}^p$, 
with $K=K_2-K_1\succ 0$ \cite[Definition 6.2]{khalil2002nonlinear}.
In this case, inequality \eqref{in.asspt4-z} 
holds with $\hat Q=-K_2^\top K_1-K_1^\top K_2$, 
$\hat S= K_1^\top+K_2^\top$ and $\hat R=-2I_n$.

\item {\it (Fully recurrent neural network)} Systems like \eqref{nonlinear.system2} with 
$B=0$ are also used to represent recurrent neural network, in which case 
$f(z)=(f(z_1)\ldots f(z_p))^\top$,  $f:\mathbb{R}\to\mathbb{R}$ is a continuously 
differential monotone nonlinear function whose derivative $f'$ is bounded from 
above on the domain $\mathbb{R}$, e.g.~$f(z_i)=\tanh(z_i)$, and $z_i= H_i^\top x+ b_i$, 
where $H_i^\top$ is the row $i$ of the matrix of weights $H$ and $b_i$ is the entry $i$ 
of the vector of biases $b$. It can be shown \cite[Section IV]{barabanov2002stability} 
that when $b=0$ (for the case case $b\ne 0$ see \cite[Section V]{barabanov2002stability}) 
the condition \eqref{in.asspt4-z} holds with $\hat Q=0$, $\hat S = \Gamma$, $\hat R =-2\Gamma$, 
where $\Gamma$ is any symmetric matrix such that for any $i=1,2,\ldots, p$, $\gamma_{ij}<0$ 
for every $j\ne i$ and $\sum_{j=1}^p\gamma_{ij}>0$ for any $i$. Hence, without loss of generality, 
$\hat R\prec 0$.  
\item {\it (Norm-bound linear difference inclusion)} 
System \eqref{nonlinear.system2} falls into 
the class of the so-called norm-bound linear difference 
inclusion \cite[Chapter 5]{boyd1994linear}.
\end{enumerate}      

\subsection{Problem formulation}

The problem of interest is to design a control law 
ensuring absolute stability for the closed loop system in the 
event that information about the system is in the form of data samples.
In this respect,
we assume to collect data of the system through offline 
experiments. We use the notation 
$U_{0},X_{0}, X_1$ and $F_{0}$ 
to denote the data matrices
\begin{subequations} \label{eq:data_matrices}
\begin{align}
& U_{0}:= 
\begin{bmatrix}
u(0) & u(1) & \cdots & u(T-1)
\end{bmatrix}\label{U0} \\
& X_{0}:= 
\begin{bmatrix}
x(0) & x(1) & \cdots & x(T-1)
\end{bmatrix}\label{X0} \\
& X_{1}:= 
\begin{bmatrix}
x(1) & x(2) & \cdots & x(T)
\end{bmatrix}\label{X1} \\
& F_0:=\begin{bmatrix}
f(0,z(0)) & f(1,z(1)) & \cdots & f(T-1,z(T-1))
\end{bmatrix}\label{F0}
\end{align}
\end{subequations}
 
We consider the following assumptions:

\begin{assume}\label{assum:measurements} 
The matrices $U_0,X_0,X_1$ and $F_0$ are known. \qedp
\end{assume}

\begin{assume}\label{assum:full-row-rank-data} 
The  matrix 
\begin{equation} \label{eq:W0}
W_0 :=
\begin{bmatrix}
U_0\\
X_{0}
\end{bmatrix} 
\end{equation}
is full row-rank. \qedp
\end{assume}

Before proceeding, we make some remarks.
 
Assumption \ref{assum:measurements} means 
that we can collect input-state samples of the system. We note in particular
that $F_0$ can be measured when the nonlinear block
$f(t,Hx)$ is physically detached from the dynamical block $x^+=Ax+Bu+Lv$, 
as schematized in Figure \ref{fig:diagram}. Besides that, 
assuming that $F_0$ is known permits us to establish a clean 
data-based analogue of absolute stability, as well as a data-based analogue 
of some related results available 
for model-based control including the circle criterion 
and the feedback Kalman-Yakubovitch-Popov Lemma.

Assumption \ref{assum:full-row-rank-data} deals instead 
with the question of richness of data. As discussed next, when this assumption
holds then it is possible to express the behavior of \eqref{nonlinear.system2} under a
control law $u=Kx$, with $K$ arbitrary, purely in terms of the data matrices in \eqref{eq:data_matrices}.
It is known that for linear controllable systems this assumption actually
reduces to a design condition that can be enforced
by suitably choosing $U_0$, see \cite{willems2005note}. 
The question of how to design experiments so as to enforce 
this condition for nonlinear systems has been recently addressed in \cite{tesi20mtns}.
Ways to relax this assumption will be discussed in the sequel.

\section{Learning control from data}\label{sec:finite-samples-nonlin}

In this section, we derive data-based conditions 
for absolute stability. The first step is to provide a data-based representation 
of the closed-loop system.
Under Assumption \ref{assum:full-row-rank-data}, for 
any matrix $K \in \mathbb R^{m \times n}$ there exists 
a matrix $G \in \mathbb R^{T \times n}$ satisfying 
\be\label{nonlinear.system.data-with H}
\begin{bmatrix}
K \\ I_n
\end{bmatrix}
= W_0G
\ee
Accordingly, the behavior of \eqref{nonlinear.system2} under a
control law $u=Kx$ can be equivalently expressed as
\be\label{nonlinear.system2-fb-data}
\setlength\arraycolsep{2pt}
\ba{rcl} 
x^+ &= & (X_1-L  F_0)G x + L v\\[0.1cm]
z &= & Hx \\[0.1cm]
v &= & f(t,z)
\ea
\ee
which follows from the chain of equalities 
\be\label{nonlinear.system.data-with H}
A+BK = \begin{bmatrix}
B & A
\end{bmatrix}
\begin{bmatrix}
K \\ I_n
\end{bmatrix}
= \begin{bmatrix}
B & A
\end{bmatrix}
\begin{bmatrix}
U_0 \\ X_0
\end{bmatrix}G
\ee
and from $X_1=AX_0+BU_0+LF_0$.

We address the absolute stabilizability problem considering 
quadratic Lyapunov functions $V(x)=x^\top P x$, in which 
case the problem becomes the one of finding two matrices $G$ and 
$P \succ 0$ such that \eqref{boh} holds
\begin{figure*}[!t]
\normalsize
{
\setcounter{MaxMatrixCols}{20}
\be\label{boh}
\begin{bmatrix}
x\\ v
\end{bmatrix}^\top
\begin{bmatrix}
G^\top (X_1-L  F_0)^\top P(X_1-L  F_0)G - P &  G^\top (X_1-L  F_0)^\top P L\\
\star & L^\top P L
\end{bmatrix}
\begin{bmatrix}
x\\ v
\end{bmatrix}<0
\ee
}
\hrulefill
\vspace*{4pt}
\end{figure*}
for all $x\ne 0$ and for all $v=f(t,z)$ that satisfy 
\be\label{in.asspt4-x}
\begin{bmatrix}
x\\  v
\end{bmatrix}^\top
\begin{bmatrix}
 Q&  S\\  S^\top &  R
\end{bmatrix}
\begin{bmatrix}
x\\  v
\end{bmatrix}
\ge 0
\ee
having defined
\be\label{QSR}
\begin{bmatrix}
 Q&  S\\  S^\top &  R
\end{bmatrix}
:=
\begin{bmatrix}
H& 0\\ 0 & I
\end{bmatrix}
^\top
\begin{bmatrix}
\hat Q& \hat S\\ \hat S^\top & \hat R
\end{bmatrix}
\begin{bmatrix}
H& 0\\ 0 & I
\end{bmatrix}
\ee 
The following result then holds. 

\begin{theorem} {\it (Data-driven absolute stabilizability)} 
\label{prop:absolute:stab}
Consider the nonlinear system \eqref{nonlinear.system2} and
let the constraint \eqref{in.asspt4-x} be {\it regular}.
Suppose that Assumption \ref{assum:measurements} and \ref{assum:full-row-rank-data} hold. 
Then, there exist two matrices $G$ and $P\succ 0$ such
that \eqref{boh} holds for all $(x,v)\ne 0$ that satisfy 
\eqref{in.asspt4-x}
\begin{enumerate}
\item ($Q\succeq  0$) 
if and only if
 there exists a $T\times n$  matrix $Y$
 such that the  matrix inequality
\be\label{lmi-nonl-stab-sample-nonl}
\begin{bmatrix}
- X_0 Y &  X_0 Y  S  & Y^\top (X_1-L F_0)^\top & X_0 Y   Q^{1/2} \\
\star &   R & L^\top & 0 \\
\star  &  \star & - X_0 Y  & 0 \\
\star  & \star  & \star   & -  I \\
\end{bmatrix}\!\!
\prec 0 
\ee
holds;
\item ($Q=0$) if and only if
there exists a  $T\times n$ matrix $Y$ such that the  matrix inequality
\be\label{lmi-nonl-stab-sample-nonl-simpler}
\begin{bmatrix}
- X_0 Y &  X_0 Y  S  & Y^\top (X_1-L F_0)^\top\\
\star &   R & L^\top\\
\star&  \star & - X_0 Y\\
\end{bmatrix}
\prec 0 
\ee
holds;
\item ($Q\preceq 0$) if there exists a  $T\times n$ matrix $Y$ 
such that the  matrix inequality \eqref{lmi-nonl-stab-sample-nonl-simpler} 
holds. In this case, the regularity of 
\eqref{in.asspt4-x} is not needed.
\end{enumerate}
In all the three cases, a state-feedback matrix
 $K$ that ensures absolute stability for the closed-loop 
 system can be computed as $K=U_0  Y (X_0Y)^{-1}$.
\end{theorem}

{\it Proof.} 1)
We want \eqref{boh} to hold 
for all $(x,v)\ne 0$ satisfying condition \eqref{in.asspt4-x}, and 
$G$ must additionally obey $X_0G=I_n$. We start by focusing 
on the relation between \eqref{boh} and  \eqref{in.asspt4-x}.
Note that 
$v=f(t,z)$ with $z=Hx$, so $v$ depends on $x$. However, this dependence can
be neglected and we can equivalently ask that \eqref{boh} holds
for all nonzero $(x,v) \in \mathbb{R}^n \times \mathbb{R}^q$ satisfying \eqref{in.asspt4-x},
where now $v$ is viewed as a free vector. This is because, as noted in
\cite[Section 2.1.2]{kh2004stability}, for any vector $v$ satisfying \eqref{in.asspt4-x}
there is a function $f(t,z)$ that satisfies \eqref{in.asspt4-x} and passes through that point.
Thus, we arrived at a stability condition for which the lossless $S$-procedure applies.
In particular, by \cite[Theorem 2.19]{kh2004stability} a necessary and sufficient 
condition to have \eqref{boh} fulfilled 
for all nonzero $(x,v)$ satisfying \eqref{in.asspt4-x} is that there exists a scalar 
$\tau\ge 0$ such that
\begin{figure*}[!t]
\normalsize
{
\be\label{coh}
\begin{bmatrix}
G^\top (X_1-L  F_0)^\top P(X_1-L  F_0)G - P +\tau  Q&  G^\top (X_1-L  F_0)^\top PL+\tau  S\\
\star & L^\top P L+\tau  R
\end{bmatrix}\prec 0
\ee
\hrulefill
}
\end{figure*}
\eqref{coh} holds for some $G$ and $P\succ0$, where
$G$ must additionally obey $X_0G=I_n$.
Without loss of generality\footnote{If $\tau=0$ the matrix inequality \eqref{boh} 
never holds since $L^\top P L\succeq 0$.}, let $\tau >0$, normalize 
the matrix $P$ ($P/\tau \to P$), and obtain \eqref{coh-norm}. 
\begin{figure*}[!t]
\normalsize
{
\be\label{coh-norm}
\begin{bmatrix}
G^\top (X_1-L  F_0)^\top P(X_1-L  F_0)G - P +  Q&  G^\top (X_1-L  F_0)^\top PL+ S\\
\star & L^\top P L+  R
\end{bmatrix}\prec 0
\ee
\hrulefill
}
\end{figure*}
By Schur complement, the latter is equivalent to 
\be\label{mi-in-proof-prop1}
\begin{bmatrix}
 - P +  Q&   S & G^\top (X_1-L F_0)^\top\\
 S^\top  & R & L^\top\\
(X_1-L  F_0)G & L & -P^{-1}
\end{bmatrix}\prec 0
\ee
To prevent the simultaneous presence of $P$ and $P^{-1}$, we 
factorize $Q= Q^{1/2} Q^{1/2}$, apply the Schur complement another time, 
left- and right-multiply by ${\rm block.diag}(P^{-1}, I, I, I)$, 
so as to obtain \eqref{lmi-nonl-stab-sample-nonl},
where $Y=G P^{-1}$. Note in particular that, in
view of this change of variable, the constraint $X_0G=I_n$ has
become $X_0Y=P^{-1}$. In turn, this implies that we can substitute 
$P^{-1}$ with $X_0Y$, which is the reason why the LMI
\eqref{lmi-nonl-stab-sample-nonl} only depends on $Y$.

Finally, the relation $K=U_0G$ gives $K=U_0  Y P=U_0  Y (X_0Y)^{-1}$. 

2) The arguments to prove 1)  continue to hold until the matrix inequality 
\eqref{mi-in-proof-prop1}, which now holds without the matrix $Q$ (since  $Q=0$). 
This allows us to directly arrive at the matrix inequality of reduced order 
\eqref{lmi-nonl-stab-sample-nonl-simpler}.

3) Since $Q\preceq 0$,  it is straightforward to realize that 
\eqref{mi-in-proof-prop1} is implied by 
\be\label{mi-in-proof-prop1-2}
\begin{bmatrix}
 - P&   S & G^\top (X_1-L F_0)^\top\\
 S^\top  & R & L^\top\\
(X_1-L  F_0)G & L & -P^{-1}
\end{bmatrix}\prec 0
\ee
Hence,
\eqref{lmi-nonl-stab-sample-nonl-simpler} is a sufficient condition for
\eqref{boh} to hold for all $(x,v)\ne 0$ that satisfy \eqref{in.asspt4-x}.
\qedp


\begin{rem}[Relaxing Assumption \ref{assum:full-row-rank-data}]
Theorem \ref{prop:absolute:stab} rests on the assumption that the matrix $W_0$
is full row rank. It is immediate to see that having $X_0$ full row rank is actually necessary
since, otherwise, \eqref{lmi-nonl-stab-sample-nonl} cannot have a solution 
because $X_0Y$ cannot be positive definite. In contrast, 
\eqref{lmi-nonl-stab-sample-nonl} might have a solution even when
$U_0$ is not full row rank. This happens when 
there exists a controller $K$ ensuring absolute stability that lies in the column 
space of $U_0$. In this sense, Assumption \ref{assum:measurements} 
guarantees that all possible controllers are evaluated.
\qedp
\end{rem}

\subsection{Discussion}

A few points worth of discussion are in order:

\subsubsection{Regularity of \eqref{in.asspt4-z} for sector bounded nonlinearities}
The regularity of the constraint \eqref{in.asspt4-z} is satisfied in some notable cases. 
In case the nonlinearity $f(t,z)$ is decoupled, namely, 
$f(t,z)={\rm col}(f_1(t,z_1), \ldots, f_p(t,z_p))$, and each 
component satisfy a sector bound constraint, then $K_1, K_2$ are 
diagonal matrices and regularity of \eqref{in.asspt4-z}  is guaranteed by 
the condition $K_2- K_1 \succ 0$, that is the interior of the sectors is non empty. 

\subsubsection{Exponential stabilizability} To guarantee exponential 
convergence of the state to the origin with decay rate $0<\rho<1$, 
it is enough to replace \eqref{lmi-nonl-stab-sample-nonl} with a weak 
matrix inequality in which the block $(1,1)$ of the matrix on the left-hand 
side is replaced by $- \rho X_0 Y$. In this case, the search for a solution 
$Y$ must be preceded by a line search on $\rho$. 

\subsubsection{Data-dependent Feedback Kalman-Yakubovitch-Popov Lemma}
Theorem \ref{prop:absolute:stab} can be viewed as a data-dependent 
Feedback Kalman-Yakubovitch-Popov Lemma \cite[Section 2.7.4]{fradkov2013nonlinear}, 
meaning that it results in a data-dependent feedback design guaranteeing the well-known 
frequency domain condition of the closed-loop system.  In fact, in the proof of 
Theorem \ref{prop:absolute:stab} we have shown that the 
condition \eqref{lmi-nonl-stab-sample-nonl} is equivalent to the existence of $P\succ 0$ 
such that \eqref{coh-norm} holds. As $Q\succeq 0$, from the block $(1,1)$ of \eqref{coh-norm} 
we deduce that the matrix $(X_1-L F_0)G =A+BK$ is Schur stable, 
hence $\det({\rm e}^{i\omega} I- A-BK)\ne 0$ for all $\omega \in \mathbb{R}$ 
and by the Kalman-Yakubovich-Popov lemma for discrete-time systems 
\cite[Theorem 2]{Rantzer1996scl}, \eqref{coh-norm} implies the frequency domain condition 
\be\label{spr.freq.domain}
\small
\!\!\begin{bmatrix}
({\rm e}^{i\omega} I- A-BK)^{-1}L\\
I 
\end{bmatrix}^*
\begin{bmatrix}
Q& S\\ S^\top & R
\end{bmatrix}
\begin{bmatrix}
({\rm e}^{i\omega} I- A-BK)^{-1}L\\
I 
\end{bmatrix}
\!\!
\prec\!0
\ee
for all $\omega\in \mathbb{R}$, where $^*$ denotes the conjugate 
transpose. Hence, condition  \eqref{lmi-nonl-stab-sample-nonl} leads 
to the existence of a gain matrix $K$ such that the frequency 
condition \eqref{spr.freq.domain} holds. 
Conversely, if $\det({\rm e}^{i\omega} I- A-BK)\ne 0$ for all 
$\omega \in \mathbb{R}$ and the  matrix inequality holds, 
then by \cite[Theorem 2]{Rantzer1996scl} there exists a matrix 
$P=P^\top$ and a real number $\tau \ge 0$ such that \eqref{coh-norm} 
holds. Note however that there is no guarantee, except in special cases, 
that $P\succ 0$, and therefore \eqref{lmi-nonl-stab-sample-nonl}, 
which would require a positive definite matrix $X_0Y=P^{-1}$, 
cannot be concluded.   

\subsubsection{Passive nonlinearities} The analysis of the 
special case of passive nonlinearities, i.e. $z^\top f(z)\ge 0$ for 
all $z\in \mathbb{R}^p$, in which case $Q=0$, $S=H^\top$, $R=0$, 
is deferred to the next section.

\subsection{Continuous-time systems} One of the features of the 
data-dependent representation introduced in \cite{depersis-tesi2020tac} 
and here adopted to deal with nonlinear systems, is that it holds for both 
continuous-time and discrete-time systems thus allowing for a unified analysis 
and design framework for both classes of systems. In this subsection we 
see how Theorem \ref{prop:absolute:stab} becomes in the case of 
continuous-time systems. Besides being of interest on its own sake, 
our motivation is to have a result to be used for some illustrative examples, 
which  are more commonly found for continuous-time systems in the literature.  

We start with 
the data-dependent representation 
for continuous-time systems, given by
\be\label{nonlinear.system2.data.ct} 
\ba{rl}
\dot x = & (X_1-L  F_0)G x + L v\\
z=&Hx\\
v=& f(t,z)
\ea
\ee
and 
\be\label{X1-ct}
X_{1}= 
\begin{bmatrix}
\dot x(t_0) & \dot x(t_1) & \ldots & \dot x(t_{T-1})
\end{bmatrix}
\ee
with $t_k$, $k=0,1,\ldots, T-1$, the sampling times at which 
measurements are taken during the off-line experiment. 
We assume that $f$ satisfies the standard conditions 
for the existence and uniqueness of the solution to the feedback interconnection, 
namely piece-wise continuity in $t$ and local Lipschitz property in $z$.

As before, we focus on the existence of a matrix 
$P\succ 0$ such that \eqref{boh-ct}
holds for all for all $x\ne 0$ and for all $v=f(t,z)$
that satisfy \eqref{in.asspt4-x}
\begin{figure*}[!t]
\normalsize
{
\setcounter{MaxMatrixCols}{20}
\be\label{boh-ct}
\begin{bmatrix}
x\\ v
\end{bmatrix}^\top
\begin{bmatrix}
G^\top (X_1-L  F_0)^\top P + P(X_1-L  F_0)G  &  P L\\
\star & 0
\end{bmatrix}
\begin{bmatrix}
x\\ v
\end{bmatrix}<0
\ee
}
\hrulefill
\vspace*{4pt}
\end{figure*}
and obtain
a necessary and sufficient condition given by the existence 
of $P\succ 0$ such that \eqref{coh-ct} holds. 
\begin{figure*}[!t]
\normalsize
{
\be\label{coh-ct}
\begin{bmatrix}
G^\top (X_1-L  F_0)^\top P+ P(X_1- L F_0)G + Q&  PL+  S\\
\star & R
\end{bmatrix}\prec 0
\ee
\hrulefill
}
\end{figure*}
In case $Q \succeq 0$, 
similar manipulations return the inequality
\be\label{lmi-nonl-stab-sample-nonl-ct}
\footnotesize
\begin{bmatrix}
Y^\top (X_1-L  F_0)^\top+ (X_1-L  F_0)Y &  L+ X_0Y S & X_0Y Q^{1/2}\\
\star & R & 0\\
\star & \star & -I 
\end{bmatrix}\prec 0
\ee
and control gain $K=U_0  Y (X_0Y)^{-1}$. In the case $Q=0$, we obtain the simpler condition 
\be\label{lmi-nonl-stab-sample-nonl-simpler-ct}
\small
\begin{bmatrix}
Y^\top (X_1- L F_0)^\top+ (X_1- L F_0)Y &  L+ X_0Y S\\
\star &  R \\
\end{bmatrix}\prec 0,
\ee
which is also a sufficient condition for the data-dependent 
absolute stabilizability of the continuous-time system when 
$Q\prec 0$. Hence, to summarize:
 
\begin{theorem}{\it (Data-driven absolute stabilizability of 
continuous-time systems)} \label{prop:absolute:stab.CT}
Consider the nonlinear continuous-time system
\be\label{nonlinear.system2-ct}
\dot x= Ax + Bu + L f(t,z), \quad z=Hx
\ee
Let Assumption \ref{assum:measurements} and \ref{assum:full-row-rank-data} hold. 
Let the constraint \eqref{in.asspt4-x} be {\it regular}. 
There exists a matrix $P \succ 0$ such that \eqref{boh-ct} 
holds for all $(x,v)\ne 0$ that satisfy \eqref{in.asspt4-x}
\begin{enumerate}
\item ($Q\succeq  0$) 
if and only if there exists a  $T\times n$ matrix $Y$ such that the matrix inequality
\eqref{lmi-nonl-stab-sample-nonl-ct} holds. 
\item ($Q=0$) 
if and only if there exists a  $T\times n$ matrix $Y$ such that the 
matrix inequality \eqref{lmi-nonl-stab-sample-nonl-simpler-ct} holds. 
\item ($Q\preceq 0$) 
if there exists a  $T\times n$ matrix $Y$ such that the 
matrix inequality \eqref{lmi-nonl-stab-sample-nonl-simpler-ct} holds. 
In this case, the regularity of \eqref{in.asspt4-x} is not needed. 
\end{enumerate}
In all the three cases, the matrix $K$ that solves the 
problem is given by $K=U_0  Y (X_0Y)^{-1}$. \qedp
\end{theorem}

An important special case is that of passive nonlinearities, namely
$z^\top f(t,z)\ge 0$ for all $z$, which
corresponds to the case where $f$ belongs to the sector 
$[0,\infty]$ \cite[Definition 6.2]{khalil2002nonlinear} Passive nonlinearities
can be written in the form \eqref{in.asspt4-x} letting
$Q=0$, $S=H^\top$ and $R=0$.
Since $R=0$, this case
does not directly fall in the previous analysis. However, it is an easy matter 
to see that, in this case, a sufficient data-dependent 
condition for the absolute stabilizability via linear feedback $u=Kx$ 
of \eqref{nonlinear.system2.data.ct} amounts to
the existence of a matrix $Y$ such that 
\be\label{spr.split}
\ba{l}
X_0 Y \succ 0\\
Y^\top (X_1-L F_0)^\top+ (X_1-L  F_0)Y\prec 0\\
L+ X_0Y H^\top=0
\ea
\ee
If a solution to \eqref{spr.split} exists then
the matrix $K$ that solves the 
problem is given by $K=U_0  Y (X_0Y)^{-1}$.
In fact, recalling that 
$A+BK=(X_1-L F_0)G$, condition \eqref{spr.split} can be recognized as 
a data-dependent condition for the \emph{strict positive realness} 
\cite[Lemma 6.3]{khalil2002nonlinear}
of the closed-loop system $(H,A+BK,L)$, where the constraint
$(A+BK)^\top P+P(A+BK) \prec 0$, $P \succ 0$, is written in the equivalent form
\[ 
Y^\top (X_1-L F_0)^\top+ (X_1-L F_0) Y \prec 0
\]
introducing the change of variable $Y=GP^{-1}$, which implies the identity 
$P^{-1}=X_0 Y$ because of the constraint $X_0G=I$. Condition \eqref{spr.split},
in turn, is a sufficient condition  
for absolute stability under passive 
nonlinearities \cite[Theorem 7.1]{khalil2002nonlinear}, the so-called 
\emph{multivariable circle criterion}.

\begin{rem}
{\it (Inferring open-loop properties from data-driven design)}  
Condition \eqref{spr.split} is also the data-dependent version 
of a well-known passifiability condition \cite[Theorem 2.12]{fradkov2013nonlinear}: 
if $L$ is full column rank, there exists a feedback $u=Kx$ which makes the 
triple $(H,A+BK,L)$ state strictly passive {\em if and only if} the 
system defined by the triple $(H, A,B)$ is minimum phase and the matrix 
$HL\prec 0$. Since $H,L$ are part of our prior knowledge, the condition 
$HL\prec 0$ can be checked. Hence, if the inequality \eqref{spr.split} is feasible, 
we infer the property of the open-loop triple $(H, A,B)$ being minimum phase  
without explicitly knowing the matrices $A,B$ but rather relying on the data 
$X_0, X_1, F_0$. Using  conditions for data-driven control to infer properties 
of an open-loop system deserves further attention in future work.  \qedp
\end{rem}

\begin{example}\label{exmp1}
We introduce an example to illustrate the application of the 
results in this section. In particular, we focus on the 
condition \eqref{spr.split}. 
We consider a pre-compensated\footnote{The actual system 
without any inner loop will be considered in Example \ref{exmp2}.} 
surge subsystem of an axial compressor model, see  e.g.~\cite{Arcak2003aut} 
\be\label{exmp1-model}
\dot x= 
\begin{bmatrix}
\frac{9}{8} & - 1\\
0 & 0 
\end{bmatrix}
x
+ 
\begin{bmatrix}
0\\
1
\end{bmatrix}
u
+ 
\begin{bmatrix}
-1\\
-\beta
\end{bmatrix}
\varphi(x_1)
\ee
with $\beta>9/8$ a parameter and $\varphi$ a passive nonlinearity 
such that $z\varphi(z)\ge 0$. Specifically, 
$\varphi(z)=\frac{1}{2} z^3+ \frac{3}{2} z^2+ \frac{9}{8} z$. 
Hence, for this example, we observe that 
a precise knowledge of $L$ is not required, any estimate 
$\hat L=\alpha \begin{bmatrix} -1 & - \beta\end{bmatrix}^\top$, 
with $\alpha>0$, used in \eqref{spr.split} does not affect the outcome of the  design. 
We perform an open-loop experiment from the initial condition 
$x(0)= \begin{bmatrix} 2 & -1 \end{bmatrix}^\top$, with $\alpha=1$, $\beta=1.2$, 
under the input $u(t)=\sin t$ over the time horizon $[0,1]$ using $T=5$ 
evenly spaced sampling times, and collect the measurements in the 
matrices $U_0, X_0, X_1, F_0$:
\[
\small
\begin{array}{rl}
U_0 =& 
\begin{bmatrix}
0   & 0.2474 &   0.4794 &   0.6816  &  0.8415
\end{bmatrix}
\\[2mm]
X_0 =& 
\begin{bmatrix}
2 &  1.269 &  1.3208 &  1.5113 &  1.7451\\
-1 &  -2.993 &   -4.3724 &   -6.0225  &  -8.2189
\end{bmatrix}
\\[3mm]
X_1 =& 
\begin{bmatrix}
-21.25 &   -5.309 &   -4.6511 &  -5.9817 &  -8.1951\\
-29.4 &   -11.428 &  -12.1319 & -15.7636 & -21.2112
\end{bmatrix}
\\[3mm]
F_0 =&
\begin{bmatrix}
 12.25 &  4.8648 &  5.2547  & 6.8522 &  9.1886
\end{bmatrix}
\end{array}\]
Assumption \ref{assum:full-row-rank-data} holds. 
We replace the data matrices in \eqref{spr.split} along with 
$\hat L=\alpha \begin{bmatrix} -1&
-\beta \end{bmatrix}^\top$,
having set $\alpha = 2$ and $H=\begin{bmatrix} 1 & 0\end{bmatrix}$. 
We remark that the parameter $\alpha$ used in \eqref{spr.split} 
is different from the value used during the experiment to stress 
that the precise knowledge of $L$ is not needed. 
We solve \eqref{spr.split} with {\tt cvx} \cite{cvx} for $Y$, 
and obtain
\[
Y =
\begin{bmatrix}
  1.2922  &  1.6018\\
   -0.1923 &     1.0528\\
    0.5113  &   -0.5863\\
    0.5192 &    -1.1827\\
   -1.0316  &    0.2419
\end{bmatrix}    
\]
from which 
\[
K=U_0  Y (X_0Y)^{-1}=\begin{bmatrix}4.3339 &  -3.7435\end{bmatrix}
\]
We observe that
\[
\small
Y^\top (X_1-L  F_0)^\top+ (X_1-L  F_0)Y
=
\begin{bmatrix}
-23.6176 & -30.3340\\
  -30.3340 & -39.1227
\end{bmatrix}
\prec 0
\]
and the entries of $\alpha L+ X_0Y H^\top$ are of order $10^{-12}$, thus  
$2x^\top P((A+BK)x+L f(z))=2x^\top P(A+BK)x- 2 \alpha^{-1} x^\top H^\top L f(z) <0$ 
for all $x$, which guarantees asymptotic stability uniformly with respect to 
any passive nonlinearity $f$.  Finally, we observe that
should the nonlinearity $f$ be time-varying, 
i.e. $f(t,z)$ during the experiment and different from the 
one appearing in the dynamics when the control is applied, 
the same result of uniform asymptotic stability will continue to hold 
as long as $z^\top f(t,z)\ge 0$ during the experiment and in the closed-loop system. \qedp
\end{example}

Giving up  the knowledge about $L,H$ 
is a difficult task. In the next section we examine one possibility 
based on strengthening 
the requirement on the collected data.

\section{Relaxing some prior knowledge by strengthened data assumptions}\label{sec:relaxing}

The last example has shown the difficulty to relax the knowledge about 
the matrices $L,H$, which influence how the nonlinearity affects the dynamics 
and which state variables appear in the nonlinear function. The situation dramatically 
changes as far as $L$ is concerned if we consider a 
stronger assumption on the set of available data. We also examine how to use 
nonlinear feedback. Specifically  
we use the term $f(t,z)$, measured for all time $t$, in the design of 
the feedback control. 

As remarked in Section \ref{sec:framework}, real time knowledge of 
$f(t,z)$ is justified in those case in which the term 
$f(t,z)$ appears as 
a physically detached block whose output can be measured. In model-based absolute 
stability theory, the case in which the nonlinearity $f$ is unknown but the signal 
$f(t,z)$
is available for on-line measurements has been considered in \cite{Arcak2003aut}. 
Alternatively, the term $Lf(t,z(t))$ can originate from modeling the nonlinearity via a vector 
of known regressors $f$ and a matrix of unknown coefficients $L$, as classically done 
in nonlinear adaptive control \cite{Sastry1989tac}. This is also the point of view taken 
in recent papers that combine sparsity-promoting techniques and machine learning 
\cite{Brunton2016PNAS}. Here, however, since we are not interested in estimating the 
dynamics but directly controlling it, we do not need to assume to know the analytic 
expression of $f$.

If we can measure in real-time $f(t,z)$, 
then we can use it also in the feedback policy, along with the state $x(t)$. 
Hence, here we consider the case in which the system 
\eqref{nonlinear.system2} is controlled via the feedback
\be\label{nonlinear.feedback}
u(t)= K x(t) + M f(t,z(t)), \quad z(t)=Hx(t)
\ee
where $K,M$ are matrices to design.
Again we stress that the feedback gains  $K,M$ are to be designed without 
knowing the analytic expression of $f$ nor $A,B,L$ 
but only the real time measurements of the vector $x(t)$ and $f(t,z)$
The matrix $H$ must be known since it appears in the matrix $Q$ (see \eqref{QSR}), 
which in turn appears in the LMI conditions that we give below.

Since the matrix $F_0$ in \eqref{F0} is known, along with $X_0, X_1, U_0$, we  
take advantage of this  knowledge by revising Assumption \ref{assum:full-row-rank-data} 
as follows:
\begin{assume}\label{assum:full-row-rank-data-rev} 
The matrix 
\[
\Psi_0:= \begin{bmatrix}
X_{0}\\
F_0\\
U_0
\end{bmatrix}
\]
 is full-row rank.
\end{assume}

For any matrix $\begin{bmatrix} K & M\end{bmatrix}\in \mathbb{R}^{m\times (n+q)}$, 
we let the matrix $G=\begin{bmatrix} G_1 & G_2\end{bmatrix}\in \mathbb{R}^{T\times (n+q)}$, 
where $G_1$ has $n$ columns and $G_2$ $q$ columns,  satisfy
\be\label{key-identity}
\begin{bmatrix}
I_n & 0_{n \times q} \\ 
0_{q \times n} & I_q \\
K & M 
\end{bmatrix}
= \begin{bmatrix}
X_{0}\\
F_0\\
U_0
\end{bmatrix}
\begin{bmatrix} G_1 & G_2\end{bmatrix}
\ee
Then we obtain the relation
\[\ba{rl}
\begin{bmatrix} A & L\end{bmatrix} + B \begin{bmatrix} K & M\end{bmatrix}
= &\left[\ba{cc|c} A & L & B\ea\right]
\begin{bmatrix}
I_n & 0_{n \times q} \\ 
0_{q \times n} & I_q \\
K & M 
\end{bmatrix}
\\[6mm]
=& 
\left[\ba{cc|c} A & L & B\ea\right]
\begin{bmatrix}X_0 \\ F_0 \\ \hline U_0
\end{bmatrix}
\begin{bmatrix} G_1 & G_2\end{bmatrix}
\\
=& X_1 \begin{bmatrix} G_1 & G_2\end{bmatrix}
\ea
\]
where we have exploited the identity $X_1=A X_0 + BU_0 + L F_0$.  
We conclude that system 
\eqref{nonlinear.system2} in closed-loop with the nonlinear feedback 
\eqref{nonlinear.feedback} is equivalent to the nonlinear data-dependent system 
\be\label{data-based-repr-nonl2}
\ba{rl}
x^+ = & X_1 \begin{bmatrix} G_1 & G_2\end{bmatrix}\begin{bmatrix} x \\
f(t,z)\end{bmatrix} \\[0.3cm] 
= & X_1 G_1 x + X_1 G_2 f(t,z) \\ 
z = & Hx
\ea
\ee
with matrices $G_1, G_2$ that satisfy \eqref{key-identity}.
We now study the absolute stability of such data-dependent system 
under the quadratic constraint assumption. For the sake of brevity, 
we only state the result in the case $Q\succeq  0$. 
The other cases are immediately obtained. 
As before, we  address the problem considering quadratic Lyapunov 
functions $V(x)=x^\top P x$, so that the problem becomes the one of 
the existence of a symmetric positive definite matrix $P$ such that 
\eqref{boh-3} holds (cf.~\eqref{boh}).
\begin{figure*}[!t]
\normalsize
{
\setcounter{MaxMatrixCols}{20}
\be\label{boh-3}
\begin{bmatrix}
x\\ v
\end{bmatrix}^\top
\begin{bmatrix}
G_1^\top X_1^\top PX_1 G_1 - P &  G_1^\top X_1^\top  P X_1 G_2\\
\star & G_2^\top X_1^\top  P X_1 G_2
\end{bmatrix}
\begin{bmatrix}
x\\ v
\end{bmatrix}<0
\ee
}
\hrulefill
\vspace*{4pt}
\end{figure*}

\begin{theorem} {\it (Data-driven absolute stabilizability II)} 
\label{prop:absolute:stab:II}
Consider the nonlinear system \eqref{nonlinear.system2}. 
Let Assumptions \ref{assum:measurements} and \ref{assum:full-row-rank-data-rev} hold and let 
$Q\succeq  0$. Let the constraint \eqref{in.asspt4-z} be {\it regular}. 
There exists a matrix $P \succ 0$ such that \eqref{boh-3} 
holds for all $x\ne 0$ and for all $v=f(t,z)$
that satisfy \eqref{in.asspt4-x}  
if and only if there exist $T\times n$, $T\times q$ 
and $n\times n$ matrices $Y_1$, $Y_2$, $W$   such that the  conditions
\be\label{lmi-nonl-stab-sample-nonl-revised}
\ba{rl}
\begin{bmatrix}
- W  & W S  & Y_1^\top X_1^\top & W  Q^{1/2} \\
\star &   R & Y_2^\top X_1^\top & 0 \\
\star  & \star & - W  & 0 \\
\star  & \star  & \star   & -I_n \\
\end{bmatrix}
&\prec 0\\[8mm]
\begin{bmatrix}
X_0 Y_1-W& X_0 Y_2\\
F_0 Y_1 &  F_0 Y_2-I_{q}
\end{bmatrix}
&= 
0
\ea
\ee
hold. In this case, the  matrices $K, M$ that solve 
the problem are given by $K=U_0 Y_1 W^{-1}$ and $M= U_0 Y_2$. 
\end{theorem}

{\it Proof.} Repeating the same 
analysis as in the proof of Theorem \ref{prop:absolute:stab} 
but this time for the representation \eqref{data-based-repr-nonl2}, 
we obtain the counterpart of \eqref{coh}, which is
\be\label{coh2}
\begin{bmatrix}
G_1^\top X_1^\top P X_1 G_1 - P +  Q&  G_1^\top X_1^\top X_1 G_2 +  S\\[2mm]
\star & G_2^\top X_1^\top  P X_1 G_2+  R
\end{bmatrix}\prec 0
\ee
where $P\succ 0$ is to be determined, and we have carried out 
the normalization $\frac{P}{\tau}\to P$. 
The same manipulations that followed  \eqref{coh} lead in this case to 
\[
\begin{bmatrix}
 - P &   S & G_1^\top X_1^\top & Q^{1/2} \\
   S^\top  &   R & G_2^\top X_1^\top & 0\\
\star & \star & -P^{-1} & 0 \\
\star & \star & \star  & -I_p \\
\end{bmatrix}\prec 0
\]
By pre- and post-multiplying the
matrix above by the matrix ${\rm block.diag}(P^{-1}, I, I, I)$ we obtain
\eqref{lmi-nonl-stab-sample-nonl-revised} having set $W:=P^{-1}$
$Y_1 := G_1 P^{-1} $, $Y_2 := G_2$.
Isolating the equation 
\be\label{sub-key-identity}
\begin{bmatrix}
I_n & 0_{n \times q} \\ 
0_{q \times n} & I_q 
\end{bmatrix}
= \begin{bmatrix}
X_{0}\\
F_0
\end{bmatrix}
\begin{bmatrix} G_1 & G_2\end{bmatrix}
\ee
in \eqref{key-identity}, taking its transpose
and multiplying it on the left by ${\rm block.diag}(P^{-1},I_p)$, we obtain 
\[\ba{rl}
\begin{bmatrix}
P^{-1} & 0 \\
0 & I_p
\end{bmatrix}
= &
\begin{bmatrix}
P^{-1} G_1^\top X_0^\top & P^{-1} G_1^\top F_0^\top \\
G_2^\top X_0^\top &  G_2^\top F_0^\top
\end{bmatrix}\\[3mm]
= &
\begin{bmatrix}
Y_1^\top X_0^\top & Y_1^\top F_0^\top \\
Y_2^\top X_0^\top &  Y_2^\top F_0^\top
\end{bmatrix}
\ea\]
that is, the constraints \eqref{key-identity} expressed in the variables $Y_1$, $Y_2$, $W$. 
In particular, since $P^{-1}= X_0 Y_1$, we have $X_0 Y_1 \succ 0$. 
Moreover, by $\begin{bmatrix} K & M \end{bmatrix}=U_0 
\begin{bmatrix} G_1 & G_2\end{bmatrix}$, we obtain
$K=U_0 G_1= U_0 Y_1 P= U_0 Y_1 (X_0 Y_1)^{-1}$ and $M= U_0 Y_2$. 
\qedp

\begin{example}\label{exmp2}
We consider a slightly revised version of Example \ref{exmp1} given by 
\[
\dot x= 
\begin{bmatrix}
\frac{9}{8} & - 1\\
0 & 0 
\end{bmatrix}
x
+ 
\begin{bmatrix}
0\\
1
\end{bmatrix}
u
+ 
\begin{bmatrix}
-1\\
0
\end{bmatrix}
\varphi(x_1)
\]
where the nonlinearity $\varphi(x_1)$ is defined as before. 
The condition \eqref{lmi-nonl-stab-sample-nonl-revised} in 
the case of passive nonlinearities for continuous-time systems is 
obtained via straighforward modifications of  \eqref{spr.split}, and 
return the following condition: there exist $T\times n$ and $T\times q$ 
matrices $Y_1$, $Y_2$   such that
\be\label{spr.split.2}
\ba{l}
Y_1^\top  X_1^\top+  X_1 Y_1\prec 0\\
X_1Y_2 + X_0 Y_1 H^\top=0\\
X_0 Y_1 \succ 0\\
X_0 Y_2 = 0 \\
F_0 Y_2 =I_p \\
F_0 Y_1 = 0 
\ea
\ee
 

We consider the same experiment as in Example \ref{exmp1}: initial condition
$x(0)= \begin{bmatrix} 2 & -1 \end{bmatrix}^\top$,  
$\alpha=2$, and input $u(t)=\sin t$ over the time horizon $[0,1]$. 
We take $T=10$ evenly spaced sampling times.  We collect the measurements 
in the matrices $U_0, X_0, X_1, F_0$, which we do not report here for 
the sake of brevity. It can be checked that 
Assumption \ref{assum:full-row-rank-data-rev}  is satisfied.  We obtain the solution
\[
\left[\ba{l|l}
Y_1 & Y_2
\ea
\right]
=
\left[\ba{rr|r}
	0.9823  & -3.5073  & -5.6005\\
   -2.0064 &    8.5180  &  12.8729\\
   -1.3370  &   7.1478 &    10.2375\\
    0.41465  &   3.1658  &   2.6801\\
    2.2302 &   -0.2915 &   -4.4256\\
    3.5496 &   -3.1425 &   -10.2866\\
    3.7054 &   -4.8273  &  -12.6223\\
    2.2325 &   -4.4031 &   -9.1124\\
   -0.7529  &  -1.5849 &   -0.4900\\
   -6.3569 &    3.4286 &   16.4407
   \ea
   \right]
\]
from which we compute the feedback gains 
\[
K=\begin{bmatrix} 7.0779   & -3.9230 \end{bmatrix}, \quad M= -3.5130
\]
and the Lyapunov matrix 
\[
P=(X_0 Y_1)^{-1}=
\begin{bmatrix} 
4.1628 &  -2.0853\\
-2.0853  &  1.1872
\end{bmatrix}   
\]
which satisfies the Lyapunov inequality 
\[
\small
Y_1^\top X_1^\top+ X_1 Y_1
=
\begin{bmatrix}
-2.5259 & -2.6865\\
 -2.6865 & -5.2943
\end{bmatrix}
\prec 0
\]
and the condition $P (L+BM)= (X_0 Y_1)^{-1} X_1 Y_2= -H^\top$. 
We observe that the program \eqref{spr.split.2} is able to correctly 
compute from data that the gain $M$ satisfies $M<-9/8$, which is a 
necessary condition for feedback \eqref{nonlinear.feedback} to render 
the closed-loop system strictly positive real \cite[Example 1]{Arcak2003aut}.
\qedp
\end{example}
 
\begin{rem}
Identities \eqref{key-identity}
suggest a  way to renounce to the knowledge of $L$ without resorting to a 
nonlinear feedback involving $f(t,z)$. 
This can be achieved by imposing $M=0$ in \eqref{key-identity}, 
which amounts to adding the constraint $0=U_0 G_2$ to 
\eqref{lmi-nonl-stab-sample-nonl-revised}. Under such conditions,
we conclude that Theorem \ref{prop:absolute:stab:II} holds  
when the feedback  is the {\em linear} one $u=Kx$. 
With respect to the case where $L$ is known,
the price to pay is that we need Assumption \ref{assum:full-row-rank-data-rev}
instead of Assumption \ref{assum:full-row-rank-data}, which is less stringent.
 \qedp
\end{rem}

\begin{example}
To illustrate the previous remark, we consider Example \ref{exmp1} 
again,\footnote{We do not use the system in Example \ref{exmp2} because 
it cannot be stabilized by a linear feedback \cite{Arcak2003aut}.} this time 
however without assuming that the matrix $L$ in \eqref{exmp1-model} is known. 
In fact, differently from Example \ref{exmp1} where we employed \eqref{spr.split}, 
here we solve \eqref{spr.split.2} with the addition of $0=U_0 G_2$. We use the 
same data $X_0, X_1, U_0, F_0$ as in Example \ref{exmp1}. 
We observe that $\begin{bmatrix}
X_{0}^\top   & 
F_0^\top  & U_0^\top
\end{bmatrix}$ is full row rank. We obtain 
\[
K=U_0  Y (X_0Y)^{-1}=\begin{bmatrix}35.8066 &  -2.1645\end{bmatrix}
\]
which makes the closed-loop matrix $A+BK$ Hurwitz, with Lyapunov matrix 
\[
P=(X_0 Y_1)^{-1}=
\begin{bmatrix} 
0.5217 &  -0.0181\\
 -0.0181  &  0.015
\end{bmatrix}   
\]
which satisfies $PL+H^\top =0$. 
\qedp
\end{example}

\section{Conclusions}\label{sec:concl}

We have presented a purely data-driven solution to derive a state feedback 
controller to stabilize 
systems with quadratic nonlinearities,
providing necessary and sufficient conditions for the absolute stabilizability 
of the closed loop system. We have discussed several variants of the results 
under different feedback (linear and nonlinear) and strengthened conditions on 
the data used for the design.
To focus on the impact of the quadratic nonlinearity in the data-dependent 
control design, we considered noiseless data. The addition of noise should be 
considered in future analysis.

\bibliographystyle{unsrt}
\bibliography{biblio-data-klein}

\end{document}